# Theoretical Investigation of Anomalous Hall and Nernst Responses in Potassium Tri-vanadium Pentantimonide

*Partha Goswami*
*Physics Dept., D.B. College (University of Delhi), Kalkaji, New Delhi-110019, India*
*Email: physicsgoswami@gmail.com*

**Abstract** We present a theoretical study of the anomalous Nernst and Hall conductance in the Kagome metal potassium tri-vanadium pent-antimonide, based on a system Hamiltonian incorporating nearest-neighbour and complex next-nearest-neighbour hopping, Rashba spin-orbit coupling, an exchange field induced by magnetic proximity, and a charge density wave potential. Our analysis reveals that the Nernst conductivity exhibits a non-monotonic temperature dependence: it increases with temperature, reaches a pronounced peak, and subsequently declines at higher temperatures due to thermal broadening, which diminishes the influence of Berry curvature. Notably, small shifts in the chemical potential can lead to dramatic changes in the Nernst signal—enhancing its magnitude or even reversing its sign—highlighting the system's sensitivity to carrier density. We further explore the anomalous Hall behaviour within this framework. The band structure hosts multiple bands with nonzero Berry curvature, and preliminary Chern number calculations suggest weak topological features: while not fully quantized, the system exhibits significant Berry curvature accumulation. Upon introducing momentum-space winding—implemented via a momentum-dependent phase in the complex hopping terms to mimic orbital magnetic flux—we observe that two bands acquire opposite Chern numbers. The remaining bands remain topologically trivial, but the system as a whole is no longer topologically inert.

## 1. Introduction

The vanadium-based Kagome metals $AV_3Sb_5$ (A = Cs, Rb, K) offer a compelling platform for investigating the intricate interplay between unconventional charge density wave (CDW) transition, chirality, nontrivial topological band structures, spin-orbit coupling (SOC), broken inversion symmetry (IS) and time-reversal symmetry (TRS) **[1–21]**. This rich confluence gives rise to a variety of exotic quantum phenomena, including anomalous Hall effect (AHE), and anomalous Nernst effect (ANE). These emergent states raise fundamental questions about the roles of the ingredients listed above in shaping the underlying electronic landscape of the Kagome metals. This study centers on potassium tri-vanadium penta-antimonide ($KV_3Sb_5$), a captivating quantum material that has emerged at the forefront of condensed matter research. Its prominence stems from the intricate interplay of the electron–phonon (e-p) coupling, Fermi surface (FS) nesting, topological band structures, and symmetry-breaking phenomena, positioning it as a fertile ground for uncovering novel quantum states and mechanisms **[1-5, 11–21]**. The e–p coupling plays a dominant role in driving the CDW transition **[17]**. In fact, the geometry of FS facilitates CDW formation through pronounced electron-lattice coupling **[18]**.

Structurally, $KV_3Sb_5$ is a layered compound in which vanadium atoms form a Kagome lattice—a two-dimensional network of corner-sharing triangles (Figure 1(a))—interspersed with potassium (K) and antimony (Sb) atoms. The crystal structure of $KV_3Sb_5$ adopts the P6/mmm space group, characterized by alternating V–Sb layers separated by K atoms. Within this framework, the vanadium sublattice exhibits a geometrically ideal Kagome configuration (Figure 1(b)). As illustrated in Figure 1(c), the antimony atoms are organized into two distinct sublattices: Sb1 forms a simple hexagonal net centered on each Kagome hexagon, while Sb2 generates graphene-like sheets positioned above and below the Kagome layers. The nearest-neighbor bond lengths are 2.74

Å for both V–V and V–Sb, and 3.16 Å for Sb–Sb. $KV_3Sb_5$ also displays remarkable electronic properties. Its band structure features Dirac cones, van Hove singularities, and flat bands—hallmarks of Kagome lattice geometry. Below approximately 78 K, the material undergoes a CDW transition, and at temperatures below ~1 K, it exhibits superconductivity. At low temperatures, $KV_3Sb_5$ enters a CDW phase which is not a conventional CDW; it exhibits spontaneous mirror symmetry breaking, manifesting in left- and right-handed configurations. Therefore, it is believed to be chiral in nature, potentially forming loop current patterns within the charge order. Its coupling with topological electronic bands opens pathways to unconventional pairing mechanisms, potentially giving rise to chiral superconductivity (SC) and the emergence of Majorana edge modes within vortex cores. The time-reversal symmetry breaking (TRSB) is induced by chiral SC and it carries profound implications for quantum computing applications.

Theoretical models [13, 19-21] suggest that loop currents can break both IS and TRS, resulting in a chiral CDW state devoid of magnetic moments. The corresponding order parameter is complex,

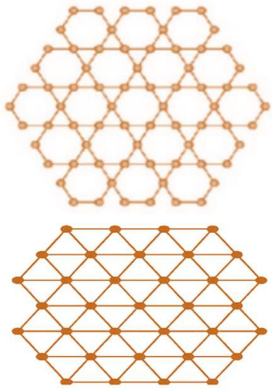

(a)

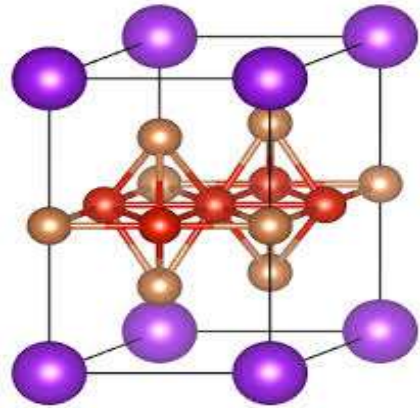

(b)

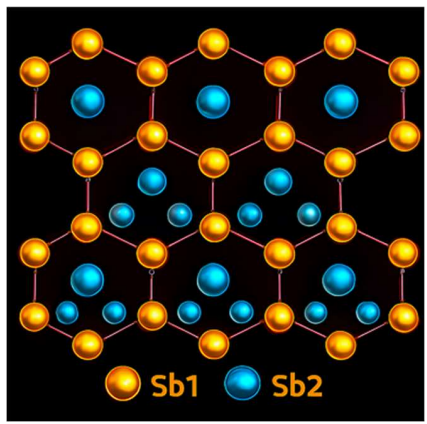

(c)

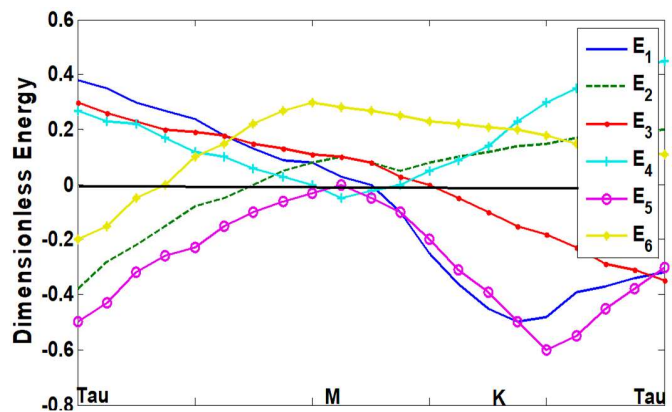

**Momentum-path (Γ- M – K – Γ)**

(d)

**Fig.1. (a)** A visual representation of non-chiral Kagome lattice and two-dimensional network of corner-sharing triangles. **(b)** The KV$_3$Sb$_5$ prototype structure (P6/mmm space group) is shown featuring a layered arrangement of V (red spheres)-Sb sheets (brown (Sb1) and orange (Sb2) spheres) separated by K (purple spheres), with the vanadium sublattice forming a structurally perfect Kagome lattice. Whereas Sb1 atoms, located within the vanadium layer or closely bonded to it, Sb2 atoms are positioned above and below the vanadium layer, forming a sandwich-like structure. **(c)** Here two distinct Sb sublattices are present with sleek metallic gradients. While the Sb1 atoms, gleaming with golden sheen, are forming a simple hexagonal net centred on each Kagome hexagon, the Sb2 atoms, shimmering in cool turquoise-blue, are creating a graphene-like Sb sheet above and below each Kagome layer. **(d)** The plot offers a conceptual illustration of the anticipated six-band electronic structure, constructed to highlight the salient features of the system. While not derived from direct numerical computation from the Hamiltonian presented below, the rendering integrates key physical ingredients and informed approximations to capture the essential characteristics. It serves as a visually guided representation, emphasizing the topological features relevant to the discussion. The plot shows spin-split flat bands near the Fermi level (represented by the horizontal solid line). The saddle point near the M point corresponds to the regions where the band curvature changes, leading to van Hove singularities and enhanced electronic interactions. The almost linear energy dispersion (Dirac cones) near the M point in the Brillouin zone is visible in the plot. The ingredients are a standard Kagome unit cell with three sublattices (A, B, C) and expanded to spinful basis (↑, ↓) to form a 6×6 minimal Bloch Hamiltonian below. The nearest neighbour (NN) plus complex next-nearest neighbour (NNN) terms, Rashba spin-orbit coupling (RSOC), and charge density wave (CDW) parameter were implemented in a compact, phenomenological way.

incorporating significant imaginary components. The orbital-selective currents lead to AHE and ANE, even in the absence of net magnetization. There is compelling evidence for broken TRS and/or IS in KV$_3$Sb$_5$[15]. Scanning tunneling microscopy reveals bond-order modulations with rotational asymmetry in KV$_3$Sb$_5$, indicative of bond-centered CDW patterns that violate the sixfold (C$_6$) rotational symmetry [22].

The central aim of this study is to provide a theoretical investigation of AHE and ANE in KV$_3$Sb$_5$. We begin with a canonical kagome unit cell comprising three sublattices (A, B, C), which is extended to a spinful basis (↑, ↓), yielding a minimal 6×6 Bloch Hamiltonian. This model incorporates nearest-neighbor (NN) hopping, complex next-nearest-neighbor (NNN) terms, Rashba spin-orbit coupling (RSOC), and a charge density wave (CDW) parameter, all implemented within a compact phenomenological framework. An illustrative rendering of the resulting six-band structure is presented in Fig. 1d. The diagram reveals spin-split flat bands proximate to the Fermi level (depicted by the horizontal solid line), alongside van Hove singularities near the M and K points—manifesting as peaks in the density of states due to enhanced band curvature. Notably, the plot exhibits nearly linear energy dispersions (Dirac cones) at these high-symmetry points in the Brillouin zone, arising from the combined effects of CDW formation and spin-orbit coupling. The Dirac bands display clear asymmetry, indicating a lifting of degeneracy and the emergence of distinct energy gaps between split bands. For parameter values for NN, NNN hopping amplitudes (∼100-250 meV), RSOC strength (∼ 50-80 meV) and CDW amplitude (∼20-25 meV) in KV$_3$Sb$_5$, we have generally referred to detailed DFT and DFT+SOC computational studies that include these quantities explicitly [23]. The Rashba spin–orbit coupling (RSOC) strength in KV$_3$Sb$_5$ is highly sensitive to surface termination, a consequence of its layered Kagome lattice structure. Angle-resolved photoemission spectroscopy (ARPES) and scanning tunneling microscopy (STM) studies in KV$_3$Sb$_5$ have revealed that Sb-terminated surfaces exhibit significantly stronger Rashba splitting compared to V-terminated ones [24]. In the analysis that follows, we therefore assume an enhanced RSOC regime. This assumption is further supported by the fact that applying a perpendicular electric field—via ionic gating or field-effect transistor (FET) configurations—can amplify structural inversion asymmetry, thereby increasing RSOC. Additionally, electron or hole doping, achieved through alkali metal substitution or intercalation

(e.g., K, Rb, or Cs), can shift the Fermi level into regions of the band structure where Rashba splitting is more pronounced.

The Nernst conductivity in our system displays a distinctly non-monotonic temperature profile: it rises with increasing temperature, peaks sharply, and then diminishes at elevated temperatures. This decline stems from thermal broadening, which suppresses the Berry curvature contributions that underpin the signal. Crucially, even minute adjustments to the chemical potential can induce pronounced shifts in the Nernst response—amplifying its magnitude or inverting its sign altogether. This pronounced sensitivity underscores the pivotal role of carrier density in modulating topological transport phenomena. We also explore the anomalous Hall behavior within our framework and the potential for realizing the quantum anomalous Hall effect (QAHE) in a modified $KV_3Sb_5$ system (the possible strategies for modification to host QAHE are given below), despite the absence of a full insulating gap in the pristine compound. This theoretical investigation is motivated by insights presented in Ref. [25]. It is essential to recognize that, whereas anomalous Hall conductance (AHC) is continuous and non-quantized, quantum anomalous Hall conductance (QAHC) is quantized in integer multiples of $\left(\frac{e^2}{h}\right)$. The former is not necessarily topological nontrivial, whereas the latter is topological with a non-zero Chern number (C). The latter stems from a topologically nontrivial band structure and intrinsic magnetism. Conversely, the former arises from Berry curvature (BC), SOC, and CDW-related orbital phenomena (CDWOP). $KV_3Sb_5$ is a promising platform for realizing QAHE, owing to its Kagome lattice hosting flat bands and Dirac points—features highly sensitive to spin-orbit coupling (SOC) and magnetic ordering. The heavy Sb atoms impart strong SOC, but QAHE requires time-reversal symmetry (TRS) breaking, likely via magnetic interactions. While a definitive magnetic gap remains undetected, anomalous Hall signals, chiral charge order, and TRS-breaking signatures below the CDW transition (~80 K) suggest proximity to a topological phase. With all key ingredients in place, $KV_3Sb_5$ stands as a near-critical system awaiting precise tuning to unlock QAHE. Its realization, however, remains challenging, hindered by unresolved theoretical complexities and prevailing experimental limitations. The following factors contribute to the challenge. First, the material does not exhibit conventional magnetic ordering; any TRS breaking likely stems from CDWOP rather than spin-based magnetism [11]. Second, no definitive magnetic gap has been observed at the Dirac points—a prerequisite for QAHE. Third, material challenges persist: disorder, imperfect band isolation, and chemical potential misalignment in real samples undermine the conditions needed for a robust topological phase. These limitations collectively constrain the emergence of QAHE in $KV_3Sb_5$. To overcome these limitations, several modification strategies have been proposed, including: (a) magnetic proximity coupling (MPC) from a substrate, (b) doping with magnetic atoms such as Mn or Cr, and (c) tuning via strain or electrostatic gating. In this study, we adopt the first approach by introducing an exchange field term $J$ arising from magnetic proximity. This term explicitly violates TRS and effectively simulates an out-of-plane magnetization, thereby facilitating the theoretical investigation of Chern insulator-like behavior.

We compute the Chern numbers for each band (from lowest to highest energy) employing the Fukui-Hatsugai-Suzuki (FHS) method [26-29]. Furthermore, introducing momentum-space winding in Sect.3, mimicking an orbital magnetic flux, through the momentum-dependence of the phase of the complex hopping, we find that two bands in the multiple band system carry opposite Chern numbers. One of the two occupied bands produces a Chern number of $C \approx +1$, whereas the other produces $C \approx -1$ and, thus, these bands demonstrate topological non-triviality. The rest of the bands exhibit trivial topology, characterized by a Chern number of C ≈ 0 in each instance, but the system as a whole is no longer topologically inert. Unlike a Kagome system $Co_3Sn_2S_2$, which exhibits a quantized total Chern number and approaches QAHE realization [30–35], we find that $KV_3Sb_5$ has a net Chern number of zero. Nonetheless, QAHE may be achievable via electron or hole doping to shift the Fermi level into Berry-curved bands. Alternatively, band structure engineering through superlattices with 2D materials like TMDs offers a promising route. This is briefly addressed in Sect. 4.

In our calculations, the exchange energy $J$ is chosen to exceed NN amplitude $t$. Realizing such a regime experimentally would necessitate integration with a strongly magnetic material exhibiting high exchange strength, along with precise interface engineering to ensure effective coupling. For clarity and consistency in our graphical representations, all energy scales have been normalized by $t$, rendering them dimensionless.

The structure of the paper is as follows. Sect. 2 introduces the theoretical model for $KV_3Sb_5$, incorporating all essential components outlined earlier. Specifically, we construct the Bloch Hamiltonian matrix in momentum space for a Kagome lattice configuration comprising a single orbital, one layer, three sublattices (A, B, C), and two spin states (↑, ↓). This formulation offers a compact yet illustrative representation of the system. In Sect. 3, we derive the corresponding band structure and compute both the anomalous Hall and anomalous Nernst conductivities. The paper ends with outlining of future research directions and a brief concluding remarks in Sect. 4.

## 2. Model

A comprehensive tight-binding Hamiltonian (TBH) for $KV_3Sb_5$ [2,3] must capture the interplay of vanadium 3d and antimony 5p orbitals, spin-orbit coupling (SOC), charge density wave (CDW) formation, and interlayer hopping. The model should reflect the quasi-two-dimensional Kagome lattice geometry and extend to a three-dimensional real-space formulation, incorporating both nearest-neighbor (NN) and next-nearest-neighbor (NNN) hopping terms, with TRS-breaking effects in the latter. SOC and interlayer coupling along the crystallographic c-axis are essential to describe the stacked Kagome layers. These ingredients give rise to electronic instabilities such as Fermi surface nesting and van Hove singularities, which enhance correlations and may drive unconventional, potentially topological superconductivity.

The $KV_3$ $Sb_5$ unit cell consists of three V atoms (sublattices A, B, and C), each contributing d-orbitals, primarily $d_{xy}$, $d_{xz}$, and $d_{yz}$. These three $d$-orbitals display odd inversion symmetry and transform in a manner analogous to vectors under the point group symmetries of the Kagome lattice. This property makes them conducive to in-plane hopping between V atoms within the Kagome network. Furthermore, the treatment of Sb atoms involves the utilization of effective p-orbitals, which may be either explicitly included or integrated out. Additionally, the consideration of spin effectively doubles the degrees of freedom. It is worth noting that the orbitals $d_{z^2}$, and $d_{x^2-y^2}$ exhibit $e_g$ symmetry, a characteristic typically associated with elevated energy levels in environments resembling octahedral or trigonal prismatic structures. Consequently, the basis vector, in terms of annihilation operators, is $\Psi = (c_{A,\alpha,\sigma}, c_{B,\alpha,\sigma}, c_{C,\alpha,\sigma})^T, \sigma = (\uparrow, \downarrow)$, where each $c_{i,\alpha,\sigma}$ represents a multi-orbital vector at site $i \in (A, B, C)$. Considering nearest-neighbour (NN) hopping $t_{i,j}$ and next-nearest-neighbour (NNN) hopping $t'_{i,j}$ between orbitals

($d_{xy}$, $d_{xz}$, and $d_{yz}$) $\alpha = m, n$ on sites $i$, and $j$, the corresponding term in TBH can be written as $H_{hop} = [\sum_{\langle im,jn \rangle, \sigma=(\uparrow,\downarrow)} t_{im,jn} c^\dagger_{im\sigma} c_{jn\sigma} + \sum_{\langle\langle im,jn \rangle\rangle, \sigma=(\uparrow,\downarrow)} t'_{im,jn} e^{i\phi} c^\dagger_{im\sigma} c_{jn\sigma} + \text{h.c.}]$, where hopping respect $C_6$ rotation and reflection. The NNN hopping includes complex phase $e^{i\phi}$ to mimic loop current. We will first consider the case $\phi$ constant. Next, we will assume $\phi = \phi(k)$ – the momentum dependence mimicking an orbital magnetic flux. The TRS breaking term comes from complex $t'_{i,j}$. Another source of TRS breaking term is MPC discussed above. The corresponding term could be written as $H_{MPC} = [\sum_{i,m,\sigma=(\uparrow,\downarrow)} J c^\dagger_{im\sigma} c_{im\sigma}]$ where $J$ is the strength of the proximity-induced exchange field. It must be noted that for investigating QAHE, a boosted out of plane magnetization is crucial (Sect. 3). The intrinsic Kane-Mele (KM) type SOC terms on V sites can be expressed as $H_{KM} = i\lambda_{SO} \sum_{\langle\langle im,jn \rangle\rangle, \sigma=(\uparrow,\downarrow),\sigma'} \nu_{i,j} c^\dagger_{im\sigma} (s_z) c_{jn\sigma'}$, where $\lambda_{SO}$ denotes the KM-SOC strength (weak to moderate for V and strong for Sb),. $\nu_{i,j} = \pm 1$ encodes the chirality of the path from $j$ to $i$, and $(s_z)$ is the Pauli matrix. For each NNN pair $\langle\langle i, j \rangle\rangle$ we compute $\nu_{i,j} = \pm 1$ depending on whether the path turns left or right at the intermediate site. The Kane-Mele SOC (inversion symmetry is preserved and sublattice localization respected) is often used as a minimal model interaction for opening topological bandgaps at Dirac points like in graphene or Kagome lattice. It must be noted that $H_{KM}$ breaks spin $SU(2)$ but preserves TRS. In real materials like KV$_3$Sb$_5$, inversion symmetry is broken either explicitly through charge-density waves, surface terminations, substrate effects, or built-in fields, or spontaneously via time-reversal or spatial symmetry breaking in electronic or lattice order. So, Rashba spin-orbit coupling(RSOC) becomes important as it originates from inversion symmetry breaking at the interface. It induces spin splitting even at TRS-invariant momenta (e.g., $\Gamma$ point), and leads to non-conserved spin and helical spin textures. The RSOC Hamiltonian, in tight-binding form for Kagome, is $H_{RSOC} = i\lambda_R \sum_{\langle im,jn \rangle, \sigma=(\uparrow,\downarrow)} c^\dagger_{im\sigma} (\mathbf{s} \times \widehat{\mathbf{d}_{ij}})_z c_{jn,-\sigma} + \text{h.c.}$, where $\lambda_R$ is the RSOC strength, $\widehat{\mathbf{d}_{ij}} = (\mathbf{r}_i - \mathbf{r}_j)/|(\mathbf{r}_i - \mathbf{r}_j)|$ is the unit vector between sites, $\mathbf{s}$ corresponds to Pauli matrices, and $(\mathbf{s} \times \widehat{\mathbf{d}_{ij}})_z = (s^x \widehat{d^y_{ij}} - s^y \widehat{d^x_{ij}}) \exp(i\mathbf{k}\cdot\boldsymbol{\delta}_{ij})$. The Hamiltonian $H_{RSOC}$ breaks inversion symmetry. Here, $\mathbf{r}_i$ is the position of the sublattice site $i \in (A, B, C)$, $\boldsymbol{\delta}_{ij}$ is the vector site $j$ to $i$, and $\mathbf{k} = (k_x, k_y)$ is the crystal momentum. The pertinent inquiry is "which (Kane-Mele or RSOC) to apply in KV$_3$Sb$_5$?". The response is, if one is simulating the topological insulating gap at Dirac points in a symmetric Kagome lattice and wishes to employ a simplified, spin-conserving model to explore Berry curvature and Z$_2$ topology, one requires the utilization of $H_{KM}$. Conversely, if one is investigating spin-split bands, helical spin textures, non-centrosymmetric CDW states, surface states, or exploring the possibility of QAHE one requires the utilization of $H_{RSOC}$. As we shall see below, the successful exploration requires cranked up $\lambda_R$ (Sect. 3). In KV$_3$Sb$_5$, possibly this is requisite, due to the strong SOC resulting from heavy V and Sb atoms. Additional reasons encompass broken inversion symmetry (IS) from CDW phases and the investigation of topological features, e.g., Dirac points, and potential quantum spin Hall (QSH) behaviour.

We model the electron-electron interaction as on-site density-density repulsion, treated in mean-field approximation as $H_{CDW} = \sum_{im,\sigma} U \langle \widehat{d_{im}} \rangle c^\dagger_{im\sigma} c_{im\sigma}$, where the CDW ansatz impose modulated density as $\langle \widehat{d_{im}} \rangle = d_0 + \delta \cos(\mathbf{r}_i \cdot \mathbf{Q})$. Here the Bravais lattice vectors are $\mathbf{a}_1 = a(1,0)$ and $\mathbf{a}_2 = a\left(\frac{1}{2}, \frac{\sqrt{3}}{2}\right)$. The position $\mathbf{r}_i$ of $(A, B, C)$ relative to the unit cell origin are typically chosen as $\mathbf{r}_A = a(0,0), \mathbf{r}_B = \frac{1}{2}\mathbf{a}_1 = a\left(\frac{1}{2}, 0\right)$, and $\mathbf{r}_C = \frac{1}{2}\mathbf{a}_2 = a\left(\frac{1}{4}, \frac{\sqrt{3}}{4}\right)$. A single-$\mathbf{Q}$

CDW vector is $\left(\frac{2\pi}{3a}\right)(1,0)$, while double-$Q$ CDW vectors are $\left(\frac{2\pi}{3a}\right)(1,0)$, and $\left(\frac{2\pi}{3a}\right)\left(-\frac{1}{2},\frac{\sqrt{3}}{2}\right)$. The typical triple CDW vectors, commonly observed vectors in KV$_3$Sb$_5$, are $\boldsymbol{Q_1} = \left(\frac{2\pi}{3a}\right)(1,0)$, $\boldsymbol{Q_2} = \left(\frac{2\pi}{3a}\right)\left(-\frac{1}{2},\frac{\sqrt{3}}{2}\right)$, and $\boldsymbol{Q_3} = \left(\frac{2\pi}{3a}\right)\left(-\frac{1}{2},-\frac{\sqrt{3}}{2}\right)$. It forms a chiral representation [22] for KV$_3$Sb$_5$ under the irreducible representations of the $C_{3z}$ point group which is a subgroup of the point group $D_{6h}$. The compound KV$_3$Sb$_5$ [16] has a layered Kagome lattice with the group $D_{6h}$ in its undistorted state. But below the CDW transition temperature ~78 K, when TRS breaking is manifested leading to anomalous Hall/Nernst effects, the symmetry is lowered (C$_6$ → C$_2$) due to lattice distortions. The three modulation wave vectors ($\boldsymbol{Q_1}, \boldsymbol{Q_2}, \boldsymbol{Q_3}$) in KV$_3$Sb$_5$ are 120° apart in reciprocal space and have equal magnitudes. This is exactly what is expected from a symmetric triple-**Q** modulation, which respects the three-fold rotational symmetry of the lattice but breaks mirror symmetry, eventually leading to chirality. How? If all three modulations are in phase, the resulting CDW is non-chiral and preserves TRS. But if the modulations have relative phase shifts — say, one lags or leads the others — the superposition leads to orbital loop currents in real space circulating around the Kagome triangles that break TRS. The TRS breaking means the system distinguishes between clockwise and counterclockwise processes. The resulting charge pattern is not mirror symmetric, giving rise to chirality. Thus, the phase difference discussed above introduces a handedness to the charge pattern — a chiral CDW. The scalar triple product ($\boldsymbol{Q_1} \times \boldsymbol{Q_2}) \cdot \boldsymbol{Q_3}$) is non-zero which further emphasizes chirality in **k**-space. Once again, how? The currents circulate in a preferred direction which is encoded by the scalar triple product and the sign of the product tells the handedness: positive means right-handed orientation, negative means left-handed. This orientation cannot be changed by simple rotation—it requires reflection, which is the essence of chirality. Upon considering the three modulation wave vectors $\boldsymbol{Q_j} = (\boldsymbol{Q_1}, \boldsymbol{Q_2}, \boldsymbol{Q_3})$, the simplified expression for the CDW term is $\sum_{\alpha,m,\sigma} \Delta_{\alpha,CDW}(\boldsymbol{r_\alpha}) c^\dagger_{\alpha m\sigma} c_{\alpha m\sigma}$, where $\alpha \in (A, B, C)$, where the CDW order parameter $\Delta_{\alpha,CDW}(\boldsymbol{r_\alpha}) = \sum_{j=1}^{3} \Delta_\alpha e^{i(Q_j \cdot r_\alpha + \phi_j)}$. The order parameter here comprises both real and imaginary components. The imaginary part induces circulating currents or orbital magnetization, which are odd under time-reversal symmetry (TRS). When considering only the real component—particularly in configurations where the phases $\phi'_j s$ are equal —the resulting CDW pattern is symmetric and achiral. However, when the phases differ by fixed increments (e.g., $\phi_j = 0, 2\pi/3, 4\pi/3$), the superposition of CDWs generates a helical or corkscrew-like modulation in real space. This non-coplanar phase arrangement breaks mirror symmetry, introducing chirality. Such chirality inherently violates TRS and can give rise to effects such as the anomalous Hall effect. The interplay between the real and imaginary components of the charge density wave (CDW) order parameter in KV$_3$Sb$_5$ gives rise to chiral CDW states exhibiting Z$_3$ nematicity, wherein the system spontaneously selects one of three symmetry-equivalent lattice directions. This emergent chirality is not only intrinsic but also magnetically switchable under moderate external fields, owing to the coupling between the magnetic field and the orbital currents associated with the chiral CDW. Such coupling facilitates reversible transitions between clockwise and counterclockwise chiral configurations. The coexistence of chiral CDW order, Z$_3$ nematic symmetry breaking, and field-tunable chirality positions KV$_3$Sb$_5$ as a versatile platform for investigating a rich landscape of quantum phenomena. These include topological phases, unconventional superconductivity, and electronic symmetry breaking, where collective electronic motion can be dynamically reoriented through the application of magnetic flux. The tunability underscores the potential of KV$_3$Sb$_5$ in advancing quantum materials research.

The interlayer hopping between stacked Kagome layers along the crystallographic c-axis plays a significant role in the electronic structure, particularly for the $d_{z^2}$ orbital. This vertical coupling, if comparable in magnitude to the dominant in-plane hopping processes, must be incorporated explicitly. The corresponding contribution to the Hamiltonian is expressed as: $H_z = [\sum_{m,n\langle i,j\rangle_z \sigma} t^z_{m,n} c^\dagger_{im\sigma} c_{jn\sigma}]$, where $t^z_{m,n}$ interlayer hopping with $(m,n)$ indexing the relevant orbitals. In matrix form, it is a $N \times N$ Bloch Hamiltonian, where $N$ = Number of sublattices × Spin × Number of orbitals × Number of layers. For a structured representation of the total Hamiltonian in momentum space, one may assume the following: (a) The Kagome lattice with 3 sublattices: (A, B, C). (b) Two spin states: ↑,↓. (c) Three orbitals ($d_{xy}$, $d_{xz}$, and $d_{yz}$). (d) Two layers stacked along the c-axis. The basis set in momentum space is $\Psi_k = (c^{A1}_{k,\uparrow} c^{B1}_{k,\uparrow} c^{C1}_{k,\uparrow} c^{A1}_{k,\downarrow} c^{B1}_{k,\downarrow} c^{C1}_{k,\downarrow} \ldots \ldots c^{C2}_{k,\downarrow})^T$. Each entry accounts for sublattice (A, B, C), layer index (1 or 2), spin orientation, and orbital state. The total basis size is, therefore, $N = 36$. This enables the comprehensive study of band structure, topological features, and interaction-driven phenomena arising from the intricate coupling between layers in Kagome-based systems.

We provide below an explicit example of the Bloch Hamiltonian matrix in momentum space for a Kagome lattice of KV$_3$Sb$_5$ with one orbital, one layer, three sublattices (A, B, C), and two spin states (↑, ↓). It helps visualize the structure without being overly large as mentioned above. This setup gives a total basis size of six in $\boldsymbol{k}$-space: $\Psi_k = \begin{pmatrix} c^A_{k,\uparrow} & c^B_{k,\uparrow} & c^C_{k,\uparrow} & c^A_{k,\downarrow} & c^B_{k,\downarrow} & c^C_{k,\downarrow} \end{pmatrix}^T$. Let's consider these terms: NN hopping of amplitude $t$, NNN Hopping of amplitude $t'$, SOC of amplitude $\lambda_R$ with imaginary off-diagonal terms (only affects NNN in Kagome systems), and the simplified expression for the CDW term. This is a Hermitian foundational model and may be used to study topological phases and flat-band physics.

$$H(\boldsymbol{k}) = \begin{bmatrix} \Delta_A + h_A & u_{AB} & u^*_{CA} & 0 & v_{AB} & -v^*_{CA} \\ u^*_{AB} & \Delta_B + h_B & u_{BC} & -v^*_{AB} & 0 & v_{BC} \\ u_{CA} & u^*_{BC} & \Delta_C + h_C & v_{CA} & -v^*_{BC} & 0 \\ 0 & -v_{AB} & v^*_{CA} & \Delta_A + h_A & u_{AB} & u^*_{CA} \\ v^*_{AB} & 0 & -v_{BC} & u^*_{AB} & \Delta_B + h_B & u_{BC} \\ -v_{CA} & v^*_{BC} & 0 & u_{CA} & u^*_{BC} & \Delta_C + h_C \end{bmatrix} \quad (1)$$

These terms introduce band bending and shift Dirac/Weyl crossings depending on the symmetry of the phase. To incorporate the MPC term into the structured Hamiltonian in (1), one can treat it as an effective Zeeman-like exchange field that acts spin-selectively on each orbital site, causing a spin-dependent onsite energy shift. For spin-up, this is $+J$, and for spin-down, it is $-J$. This means $H(k) \rightarrow H(k) + H_{\text{MPC}} = H(k) + J \cdot \tau_z \otimes I_3$, where $\tau_z$ acts in spin space and $I_3$ is the identity matrix in the sublattice space. The term $H_{\text{MPC}}$ breaks TRS and lifts spin degeneracy, and when combined with SOC and Berry curvature, it enhances topological effects. Additionally, MPC does not mix spin states, so SOC and hopping terms remain unchanged. The combined Hamiltonian encapsulates the essential physics of the single-layer Kagome system, notably its intrinsic chirality, topological band inversion, and spin–orbit coupling–induced degeneracy lifting. It may also offer valuable insight into the QAHE-related considerations outlined in Sect. 1. Here, Kagome lattice has 3 sites per unit cell, include hopping between nearest neighbours (A↔B, B↔C, C↔A), SOC only couples spins on NNN paths with complex phases, and CDW introduces on-site energy variation, breaking translational symmetry but treated here in momentum space as diagonal terms. The CDW terms ($\Delta_A, \Delta_B, \Delta_C$), in their simplest form, correspond to on-site energies for the sublattices (A, B, C). The NN hopping terms for model of the metal KV$_3$Sb$_5$ are $u_{AB}(\mathbf{k}) = t(1 + e^{i\mathbf{k}\cdot\boldsymbol{\delta}_{AB}})$, $u_{BC}(\mathbf{k}) = t(1 + e^{i\mathbf{k}\cdot\boldsymbol{\delta}_{BC}})$, and $u_{CA}(\mathbf{k}) = t(1 + e^{i\mathbf{k}\cdot\boldsymbol{\delta}_{CA}})$. The exponential terms capture the hopping to the neighbour connected via vector $\boldsymbol{\delta}_{ij}$, preserving Bloch periodicity. The Kagome lattice consists of corner-sharing triangles. Given the Bravais lattice vectors are $\mathbf{a}_1 = a(1,0)$ and $\mathbf{a}_2 = a\left(\frac{1}{2}, \frac{\sqrt{3}}{2}\right)$, then what are the NN vectors are $\boldsymbol{\delta}_{AB}, \boldsymbol{\delta}_{BC}$, and $\boldsymbol{\delta}_{CA}$? We define the NN vectors so they connect the three sites of the triangle (A→B, B→C, C→A) symmetrically: $\boldsymbol{\delta}_{AB} = \left(\frac{a}{2}\right)(1,\sqrt{3})$, $\boldsymbol{\delta}_{BC} = \left(\frac{a}{2}\right)(-1,\sqrt{3})$, and $\boldsymbol{\delta}_{CA} = a(0,-1)$. The NNN hopping corresponds to $h_\alpha$'s where $\alpha \in (A, B, C)$. The NNN hopping terms can and often should be complex, especially in Kagome systems exhibiting topological features. In fact, in a typical Kagome lattice, the second-neighbour paths form closed loops (e.g. hexagons), and introducing complex phase factors (typically from orbital flux or effective magnetic fields) breaks time-reversal or inversion symmetry, enabling phenomena like topological band structures, band splitting and flat bands, and non-trivial Berry curvature. Consider now sublattice A and vectors $\boldsymbol{\delta}_i$. These vectors point to NNN sites around hexagon centred on A sites. We have $\boldsymbol{\delta}_1 = \mathbf{a}_1 + \mathbf{a}_2 = a\left(\frac{3}{2}, \frac{\sqrt{3}}{2}\right)$, $\boldsymbol{\delta}_2 = -\mathbf{a}_1 + \mathbf{a}_2 = a\left(-\frac{1}{2}, \frac{\sqrt{3}}{2}\right)$, and $\boldsymbol{\delta}_3 = -\mathbf{a}_2 = a\left(-\frac{1}{2}, -\frac{\sqrt{3}}{2}\right)$. For sublattice B, the corresponding vectors $\boldsymbol{\delta}'_i$ are given by $\boldsymbol{\delta}'_1 = a(-1,0)$, $\boldsymbol{\delta}'_2 = \left(\frac{a}{2}\right)(1,-\sqrt{3})$, and $\boldsymbol{\delta}'_3 = \left(\frac{a}{2}\right)(1,\sqrt{3})$. These are 120° rotated version of A vectors. Similarly, for the sublattice C rotated again by another 120°, the corresponding vectors $\boldsymbol{\delta}''_i$ are given by $\boldsymbol{\delta}''_1 = \left(\frac{a}{2}\right)(1,-\sqrt{3})$, $\boldsymbol{\delta}''_2 = \left(\frac{a}{2}\right)(3,\sqrt{3})$, and $\boldsymbol{\delta}''_3 = a(-1,0)$. These are NNN displacement vectors from a site on a sublattice to one of its NNN partners. Mathematically, this means NNN hopping terms take the form

$$h_A(\mathbf{k}) = t'\left[e^{i\left(\mathbf{k}\cdot\boldsymbol{\delta}_1 + \frac{\pi}{2}\right)} + e^{i\left(\mathbf{k}\cdot\boldsymbol{\delta}_2 - \frac{\pi}{2}\right)} + e^{i\left(\mathbf{k}\cdot\boldsymbol{\delta}_3 + \frac{\pi}{2}\right)}\right] - \mu,$$

$$h_B(\mathbf{k}) = t'\left[e^{i\left(\mathbf{k}\cdot\boldsymbol{\delta}'_1 - \frac{\pi}{2}\right)} + e^{i\left(\mathbf{k}\cdot\boldsymbol{\delta}'_2 + \frac{\pi}{2}\right)} + e^{i\left(\mathbf{k}\cdot\boldsymbol{\delta}'_3 - \frac{\pi}{2}\right)}\right] - \mu,$$

$$h_C(\mathbf{k}) = t'\left[e^{i\left(\mathbf{k}\cdot\boldsymbol{\delta}''_1 + \frac{\pi}{2}\right)} + e^{i\left(\mathbf{k}\cdot\boldsymbol{\delta}''_2 + \frac{\pi}{2}\right)} + e^{i\left(\mathbf{k}\cdot\boldsymbol{\delta}''_3 - \frac{\pi}{2}\right)}\right] - \mu, \qquad (2)$$

where $\mu$ is the chemical potential of the fermion number. The off-diagonal SOC terms are

$$v_{AB}(\mathbf{k}) = i\lambda_R \sin(\mathbf{k}\cdot\mathbf{b}_1), \quad v_{BC}(\mathbf{k}) = i\lambda_R \sin(\mathbf{k}\cdot\mathbf{b}_2), \quad v_{CA}(\mathbf{k}) = i\lambda_R \sin(\mathbf{k}\cdot\mathbf{b}_3). \qquad (3)$$

We have $\mathbf{b}_1 = a(1,0)$, $\mathbf{b}_2 = a\left(-\frac{1}{2}, \frac{\sqrt{3}}{2}\right)$, and $\mathbf{b}_3 = a\left(-\frac{1}{2}, -\frac{\sqrt{3}}{2}\right)$. These correspond to three equivalent directions wrapping around the hexagonal center with uniform chirality. For symmetric Rashba hopping in KV$_3$Sb$_5$'s single triangle or minimal model, three $\mathbf{b}$ vectors for NNN paths suffice when spin-flip terms are sublattice-specific. However, six $\mathbf{b}$ vectors are essential for full tight-binding simulations of Kagome lattices with RSOC due to the multiple NNN paths connecting sublattice pairs. These vectors are given by

$$\boldsymbol{b_1} = \boldsymbol{a_1} + \boldsymbol{a_2} = a\left(\frac{3}{2}, \frac{\sqrt{3}}{2}\right), \quad \boldsymbol{b_2} = -\boldsymbol{a_1} + \boldsymbol{a_2} = a\left(-\frac{1}{2}, \frac{\sqrt{3}}{2}\right), \boldsymbol{b_3} = -\boldsymbol{a_1} - \boldsymbol{a_2} = a\left(-\frac{3}{2}, -\frac{\sqrt{3}}{2}\right),$$

$$\boldsymbol{b_4} = \boldsymbol{a_1} - \boldsymbol{a_2} = a\left(\frac{1}{2}, -\frac{\sqrt{3}}{2}\right), \boldsymbol{b_5} = -\boldsymbol{a_2} = a\left(-\frac{1}{2}, -\frac{\sqrt{3}}{2}\right), \text{ and } \boldsymbol{b_6} = \boldsymbol{a_2} = a\left(\frac{1}{2}, -\frac{\sqrt{3}}{2}\right). \quad (4)$$

The Kagome lattice's structure, with three sublattices and dual NNN paths between each pair of differing chirality, necessitates six unique $\boldsymbol{b}$ vectors to accurately model RSOC's effects on angular dependence and symmetry breaking.

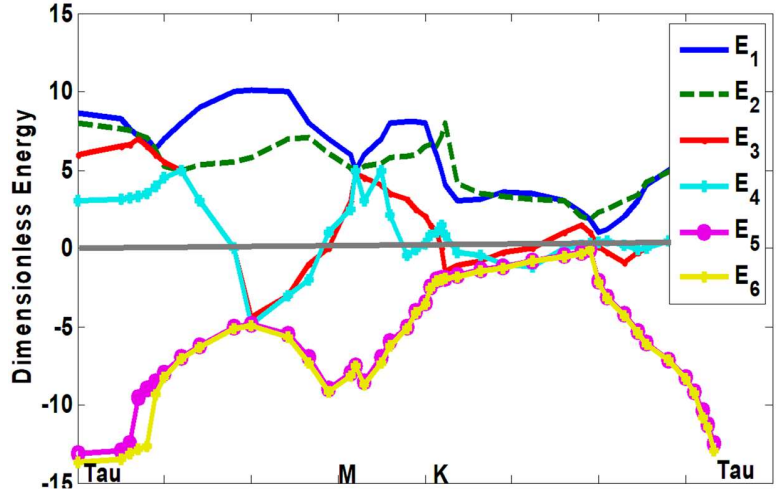

(a)

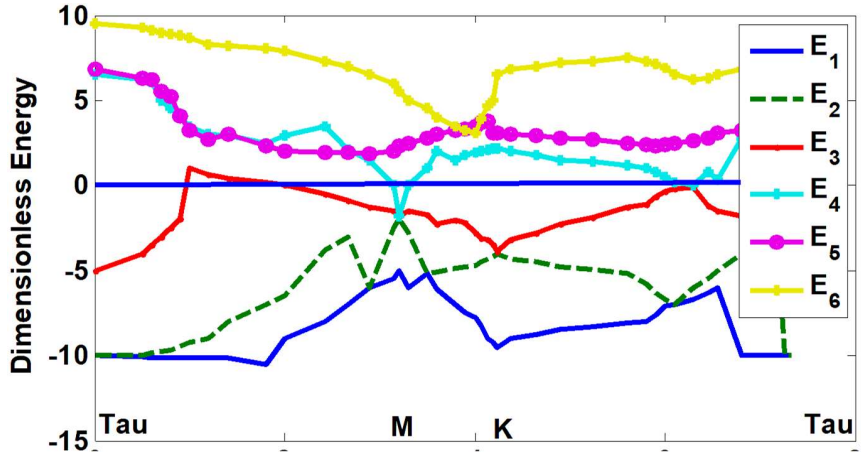

(b)

**Fig. 2. (a) (b)** The (six) band structure for the Kagome lattice model of $KV_3Sb_5$ using NN hopping, NNN hopping (complex with phase $\phi = \pi/3$ in *a*, but $\phi = \phi(\boldsymbol{k}) = (\sin(k_x) - \sin(k_y))$ in *b* ), RSOC, and CDW amplitude. The six bands

arise from the combination of three sublattices (A, B, C) and two spin states (↑, ↓). The high-symmetry path followed is $\Gamma \to M \to K \to \Gamma$. The parameter values used are $t = 1, t' = 0.44$ (in Fig. a) and $t' = 0.86$ (in Fig. b), $\lambda_R = 0.14$ (in Fig. a) and $\lambda_R = 0.80$ (in Fig. b), $\Delta_{CDW} = 0.14$, and $J = 0.71$ (in Fig. a), and $J = 2.3$ (in Fig. b). The graphical representation exhibits the spin splitting due to SOC. The spectral gaps are induced by CDW. The figure exhibits a band structure with saddle points, leading to Berry curvature hotspots—key ingredients for non-zero Chern numbers. There is near-flat band feature close to the Fermi level.

## 3. Method

### A. Band Structure

Our model of KV$_3$Sb$_5$ incorporates several essential physical mechanisms: NN hopping, complex NNN hopping with a phase factor $\phi = \pi/3$, RSOC, and a CDW amplitude. The resultant band structure, which emerges from three sublattices (A, B, C) combined with spin degrees of freedom (↑, ↓), comprises six distinct bands as in Fig.2. We need to lean on numerical analysis to obtain the eigenvalues of Eq.(1). We use the 'Matlab' package for this purpose. The numerical values of the eigenenergy and the corresponding eigenvectors are obtained by the command [V,D] = eig($H(\mathbf{k})$). The command returns diagonal matrix D of eigenvalues and matrix V whose columns are the corresponding right eigenvectors, so that $H(\mathbf{k})$ *V = V*D. For each k-point, in the chosen $k$-path, this process is repeated. The plots of the energy eigenvalues $E_j(k)$ ($j = 1,2,\ldots\ldots,6$) as, obtained in this manner, are shown in Fig. 2. The parameter values used are $t = 1, t' = 0.44$ (in Fig. a) $and$ $t' = 0.86$ (in Fig. b), $\lambda_R = 0.14$ (in Fig. a) and $\lambda_R = 0.80$ (in Fig. b), $\Delta_{CDW} = 0.14$, and $J = 0.71$ (in Fig. a), and $J = 2.3$ (in Fig.b). We have plotted the band dispersion along high-symmetry lines in the 2D Brillouin zone ($\Gamma \to K \to M \to \Gamma$). The computed band structure reveals the characteristic flat band associated with kagome systems. These flat regions are attributed to destructive interference patterns arising from the lattice geometry and are known to promote electronic instabilities. Moreover, spin splitting induced by RSOC is readily apparent, emphasizing the role of spin–orbit interactions in lifting degeneracies. Importantly, in Fig.2b, the bands $E_3$ and $E_4$ ($E_5$ and $E_6$) display band-inversion near M-point (between M and K points). The introduction of a CDW leads to spectral gaps, indicating a reconstruction of the Fermi surface and breaking of translational symmetry. Within this structure, a saddle point, near the M- point, emerges that act as sources of enhanced Berry curvature. In fact, a saddle point in a band structure refers to a special point in the energy dispersion of electrons in a crystal leading to a van Hove singularity, manifesting as a distinct peak indicative of enhanced density of states due to band curvature. The resulting band structure also reveals asymmetric Dirac bands near the K- and M- points, indicating broken inversion and time-reversal symmetry.

### B. Anomalous Conductivities

The anomalous Nernst effect (ANE)—a transverse voltage appearing even without a magnetic field, driven by Berry curvature, is a geometric feature of quantum states. Both AHE and ANE effects reveal how quantum structure—entanglement in Hilbert space or geometry in momentum space—reshapes energy transport beyond classical expectations. The exploration of anomalous Nernst effect (ANE) in KV$_3$Sb$_5$ [36,37] offers one of the most direct ways to uncover the hidden quantum and topological features of this remarkable Kagome metal. In Kagome metals such as KV$_3$Sb$_5$, Rb V$_3$Sb$_5$, and Cs V$_3$Sb$_5$, recent experiments [38] have revealed pronounced anomalous Nernst signal below approximately 100 K. This effect originate from TRS breaking within CDW phase which induces a nontrivial Berry curvature (BC) texture. While it does not exhibit the fully quantized response characteristic of a quantum anomalous Hall (QAH) insulator, it shares the same underlying topological mechanisms, manifesting them in a partially filled band structure.

We will now provide formula to calculate anomalous Nernst conductivity(ANC) $\alpha_{xy}(T)$. It needs to be computed by integrating the Berry curvatures close to the Fermi level along with the entropy density over first BZ [39], i.e. $\alpha_{xy}(\mu, T) = k_B \frac{e}{\hbar} \sum_n \int d\mathbf{k}\, \Omega_{xy}^{(n)}\, s(E_n(k))$ where the entropy density $s(E_n(k))$ is given by the expression

$$s(E_n) = \frac{E_n - \mu}{k_B T} f(E_n) + \log\left(1 + \exp\left(\frac{\mu - E_n}{k_B T}\right)\right), f(E_n) = \frac{1}{1+\exp\left(\frac{E_n-\mu}{k_B T}\right)}. \tag{5}$$

Here, $\mu$ is the chemical potential, $\Omega_{xy}^{(n)}$ is the BC in the z-direction corresponding to the energy band $E_n$, $n$ is the band index close to the Fermi level, and $f(E)$ is the Fermi-Dirac distribution with $k_B$ the Boltzmann constant. In the low-temperature limit, upon using the Mott relation [40] we obtain

$$\alpha_{xy}(\mu) \approx \frac{\frac{\pi^2}{3} k_B^2 T}{e} \sum_n \int d\mathbf{k}\, \Omega_{xy}^{(n)} \frac{\partial f(E_n(k))}{\partial E_n(k)}. \tag{6}$$

The anomalous Nernst conductivity (ANC), and the anomalous Hall conductivity (AHC) are calculated here by integrating the Berry curvature (BC) on a k-mesh-grid of the Brillouin zone. The expression of AHC, on the other hand, is $\sigma_{xy} = -\left(\frac{e^2}{\hbar}\right) \sum_n \int_{BZ} \frac{d^3k}{(2\pi)^3} f(E_n(k) - \mu)\, \Omega_n^z(k)$, where μ is the chemical potential of the fermion number, $n$ is the occupied band index, $f(E_n(k) - \mu)$ is the Fermi-Dirac distribution and $\Omega_n^z(k)$ is the z-component of BC for the $n$th occupied band. The Berry curvature could be calculated using the Kubo formula [32,41]

$$\Omega_n^z(k) = -2\hbar^2 \left[ Im \sum_{m \neq n} (E_n(k) - E_m(k))^{-2} \langle n, k | \widehat{v_x} | m, k \rangle \langle m, k | \widehat{v_y} | n, k \rangle \right]. \tag{7}$$

Here $k$ is the Bloch wave vector, $E_n(k)$ is the band energy, $|n, k\rangle$ are the Bloch functions of a single band. The operator $\hat{v}_j$ represents the velocity in the $j$ direction. We recall that for a system in a periodic potential and its Bloch states as the eigenstates, in view of the Heisenberg equation of motion $i\hbar \frac{d\hat{x}}{dt} = [\hat{x}, \hat{H}]$, the identity

$$\langle m, \mathbf{k}' | v_\alpha | n, \mathbf{k} \rangle = \left(\frac{1}{\hbar}\right) (E_n(\mathbf{k}') - E_m(\mathbf{k})) \left\langle m, \mathbf{k}' \left| \frac{\partial}{\partial k_\alpha} \right| n, \mathbf{k} \right\rangle \tag{8}$$

is satisfied. Upon using this identity, we obtain Hall conductivity in the zero temperature limit as $\sigma_{xy} = C\left(\frac{e^2}{\hbar}\right)$ where $C = \sum_n C_n$, $C_n = \int \int_{BZ} \Omega_{xy}(k) \frac{d^2k}{(2\pi)^2}$, where the integral is over the entire Brillouin zone(BZ) and corresponds to the sum of Berry curvatures of all occupied bands. If the Chern number $C$ for our system is quantized, we have the system featuring quantum anomalous Hall effect (QAHE), else it is simple anomalous Hall effect. The z-component of BC is

$$\Omega_{xy}(k) = \left(\frac{\partial A_{n,y}}{\partial k_x} - \frac{\partial A_{n,x}}{\partial k_y}\right) = -2\ Im\ \left\langle\frac{\partial \psi_{n,k}}{\partial k_x}\bigg|\frac{\partial \psi_{n,k}}{\partial k_y}\right\rangle \qquad (9)$$

where $\psi_{n,k} = |n,k\rangle$. The vector potential $A_n(\boldsymbol{k})$ is the Berry connection and $\nabla_{\boldsymbol{k}} \times A_n(\boldsymbol{k}) = \Omega_n(\boldsymbol{k})$ is derived from it. While the derivative calculations involved in the Kubo formula are conceptually straightforward, they are computationally intensive. To investigate the Chern number, we therefore adopted the Fukui-Hatsugai-Suzuki (FHS) method**[26-29]**, which assumes a discretized Brillouin zone. This technique is grounded in lattice gauge theory, replacing continuous derivatives with discrete link variables. Crucially, the FHS method preserves gauge invariance under lattice boundary conditions. The Chern number is extracted from the Wilson loop **[26-29]**, defined as the product of link variables around a plaquette in momentum space. This formulation remains invariant under local phase transformations of the Bloch wavefunctions, ensuring robust topological characterization.

A quantitative topological analysis was conducted by evaluating BC of individual bands on a coarse N × N ( N = 50) momentum-space grid using the parameter values as $t = 1$, $t' = 0.44$, $\lambda_R = 0.14$, $J = 0.71$, and $\Delta_{CDW} = 0.14$. This results in grainy BC distribution as shown in the sample in Fig. 3a. We have obtained the following values of the Chern number $C_1 = +0.00$, $C_2 = -0.02$, $C_3 = +0.03$, $C_4 = +0.09$, $C_5 = +0.01$, and $C_6 = -0.04$ corresponding to $E_1$, $E_2$, $E_3$, $E_4$, $E_5$, and $E_6$, respectively, in Fig.2a. The higher resolutions (N = 200) improve the visual smoothness of BC plots as in Figs.3b, 3c, 3d, 3e, 3f, and 3g without compromising feature like the presence of an intrinsic anomalous Hall effect (AHE). It may be noted that small non-zero Chern numbers (|C|<<1) contribute to transverse conductivity. The NNN hopping is complex with phase $\phi = \pi/3$. The results hint at possible topological transitions and potential Chern band formation with tweaking of parameters. In other words, these create a fertile ground for further exploration of topological transitions, though the current scenario has not yet crossed the threshold for integer Chern numbers.

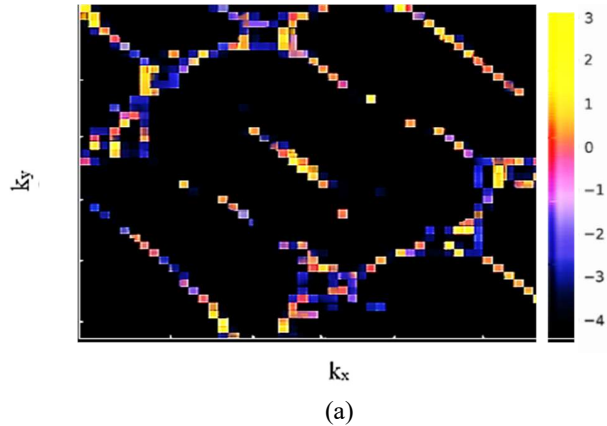

(a)

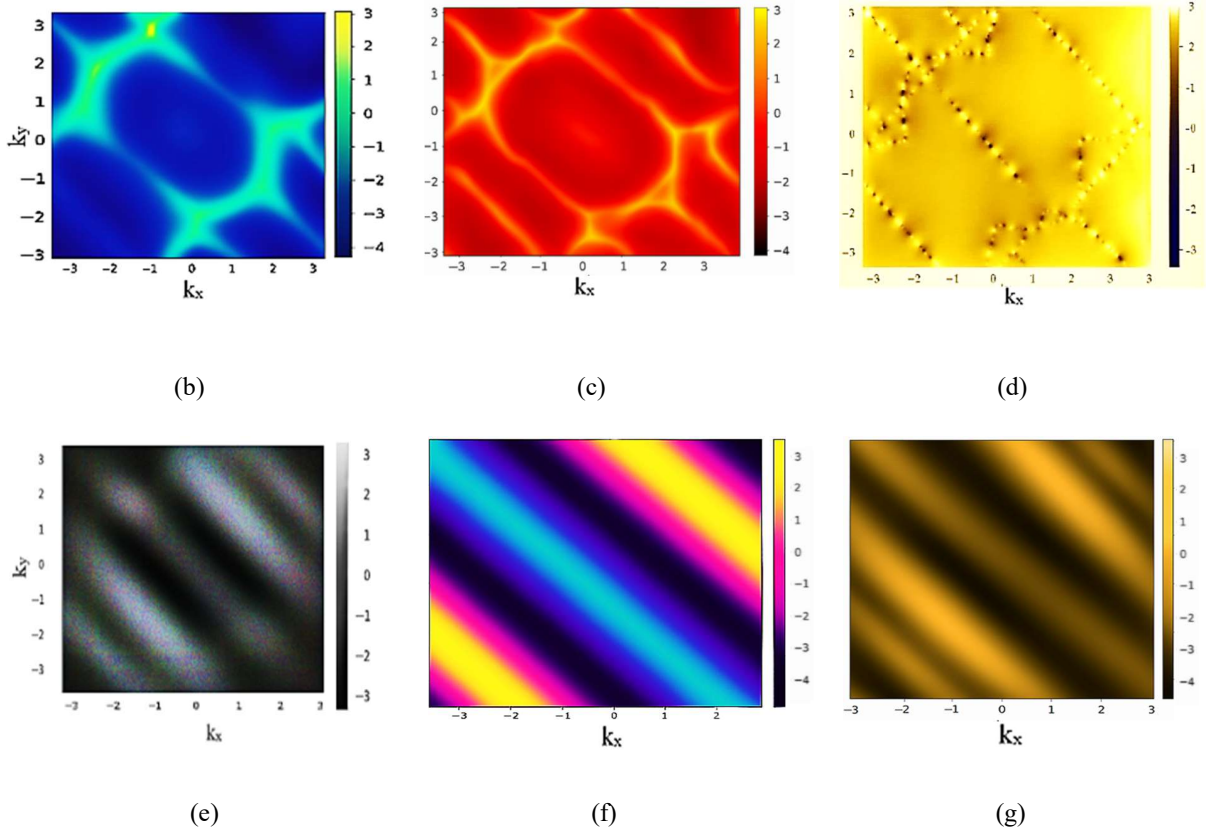

**Fig.3. (a)** The 2D plot of the Berry curvature as a function of $k_x$ and $k_y\{k_x, k_y \in (-\pi, \pi)\}$ corresponding to the energies $E_1$ in Fig.2a, with 50 × 50 resolution. **(b)-(g)** 2D plots of the Berry curvature as a function of $k_x$ and $k_y$ corresponding to the energies $E_1, E_2, E_3, E_4, E_5,$ and $E_6$ in Fig.2a, respectively, with 200 × 200 resolution. The parameter values used are $t = 1$, $t' = 0.44$, $\lambda_R = 0.14$, $J = 0.71$, and $\Delta_{CDW} = 0.14$. The NNN hopping is complex with phase $\phi = \pi/3$. We obtain the following values of the Chern number $C_1 = +0.00$, $C_2 = -0.02$, $C_3 = +0.03$, $C_4 = +0.09$, $C_5 = +0.01,$ and $C_6 = -0.04$ corresponding to $E_1, E_2, E_3, E_4, E_5,$ and $E_6$, respectively, in Fig.2a. In these figures, the color bar is fully annotated ranging from –3 at the bottom to +3 at the top; the intermediate ticks are positioned at –2, –1, 0, 1, and 2.

As regards $\alpha_{xy}$, upon assuming the phase of the complex NNN hopping as $\phi = \pi/3$, we have computed it using the numerical value of the parameters as $t = 1$, $t' = 0.86, \lambda_R = 0.8, J = 2.3, \mu = 0.01$, and $\Delta_{CDW} = 0.14$. We have presented the outcome as a 2D graphical representation of $\alpha_{xy}(T)$ as a function of $t/k_BT$ in Fig. 4a. The derivative $(-\frac{\partial f}{\partial E})$ of the Fermi-Dirac distribution is sharply peaked around the chemical potential E = μ, with a characteristic width on the order of $k_BT$. Consequently, at low temperatures, only electronic states in close proximity to μ contribute significantly to transport phenomena. Typically, the anomalous Nernst coefficient $\alpha_{xy}$ for $KV_3Sb_5$ increases with temperature, attains a maximum, and subsequently decreases at higher temperatures due to thermal smearing, which diminishes the influence of Berry curvature. This trend is qualitatively consistent with the findings reported in Ref. **[42]**.

In materials like $KV_3Sb_5$, the Kagome lattice and charge density wave create localized "hot spots" of Berry curvature near band crossings. As a result, even small variations in μ (for instance, via doping or gating) can cause substantial changes in the measured $\alpha_{xy}$ signal, sometimes flipping its sign or enhancing its magnitude significantly. In practical terms, tuning the chemical potential $\mu$ — by external means or intrinsic effects—acts as a "knob" for controlling the anomalous transport responses in topological and correlated electron systems. In order to highlight this, we have shown the dependence of $\alpha_{xy}$ as a function of chemical potential $\frac{\mu}{t}$ in Fig.4b. It typically displays pronounced peaks near band crossings where Berry curvature is large. It also shows node or, zero crossing where Berry curvature changes sign. Polynomial fitting is also employed to the plots to elucidate the underlying trends in the data.

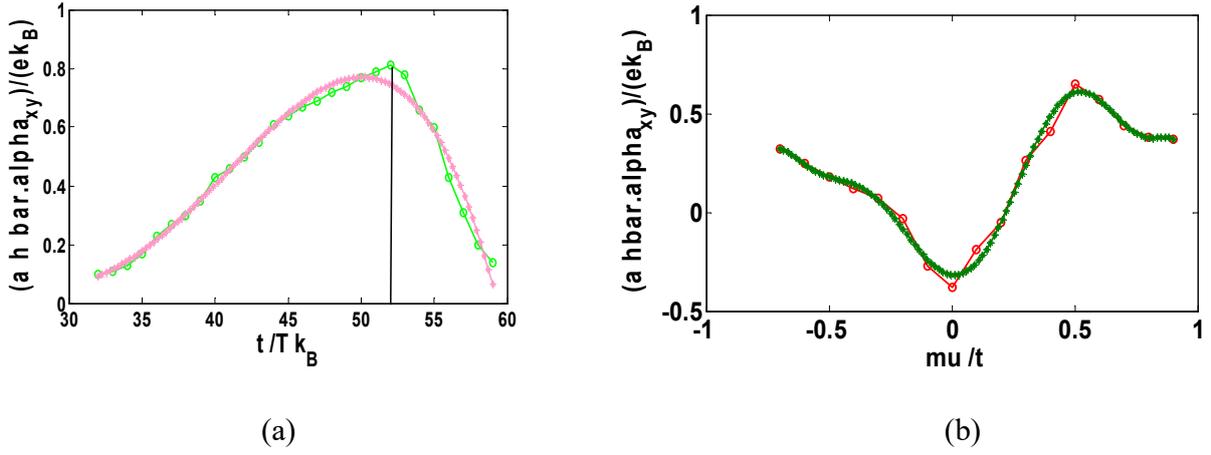

(a)                                    (b)

**Fig.4. (a)** Two-dimensional plot of the anomalous Nernst coefficient $\alpha_{xy}(T)$ are presented as a function of the dimensionless parameter $t/k_B T$. The plot shows that near $t/k_B T = 25$, $\alpha_{xy}(T)$ assumes much larger value compared to those at higher temperature in the limited temperature range considered here. The numerical value of the parameters used are $t = 1$, $t' = 0.86$, $\lambda_R = 0.8$, $J = 2.3$, $\mu = 0.01$, $\phi = \pi/3$, and $\Delta_{CDW} = 0.14$. **(b)** The dependence of $\alpha_{xy}$ as a function of chemical potential $\frac{\mu}{t}$. The numerical value of the parameters used are the same as in Fig.4a. Polynomial fitting is applied to these plots to capture the underlying trends.

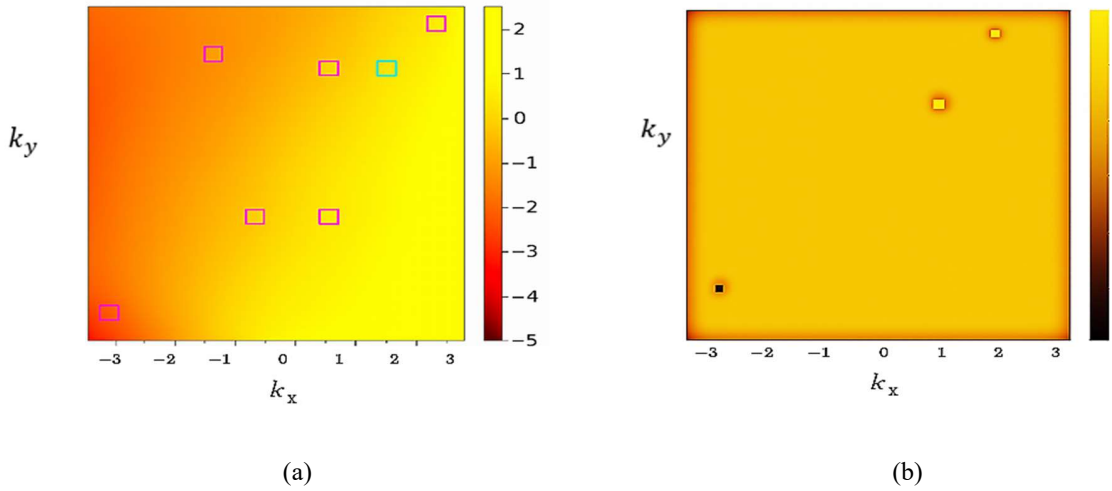

(a)                                    (b)

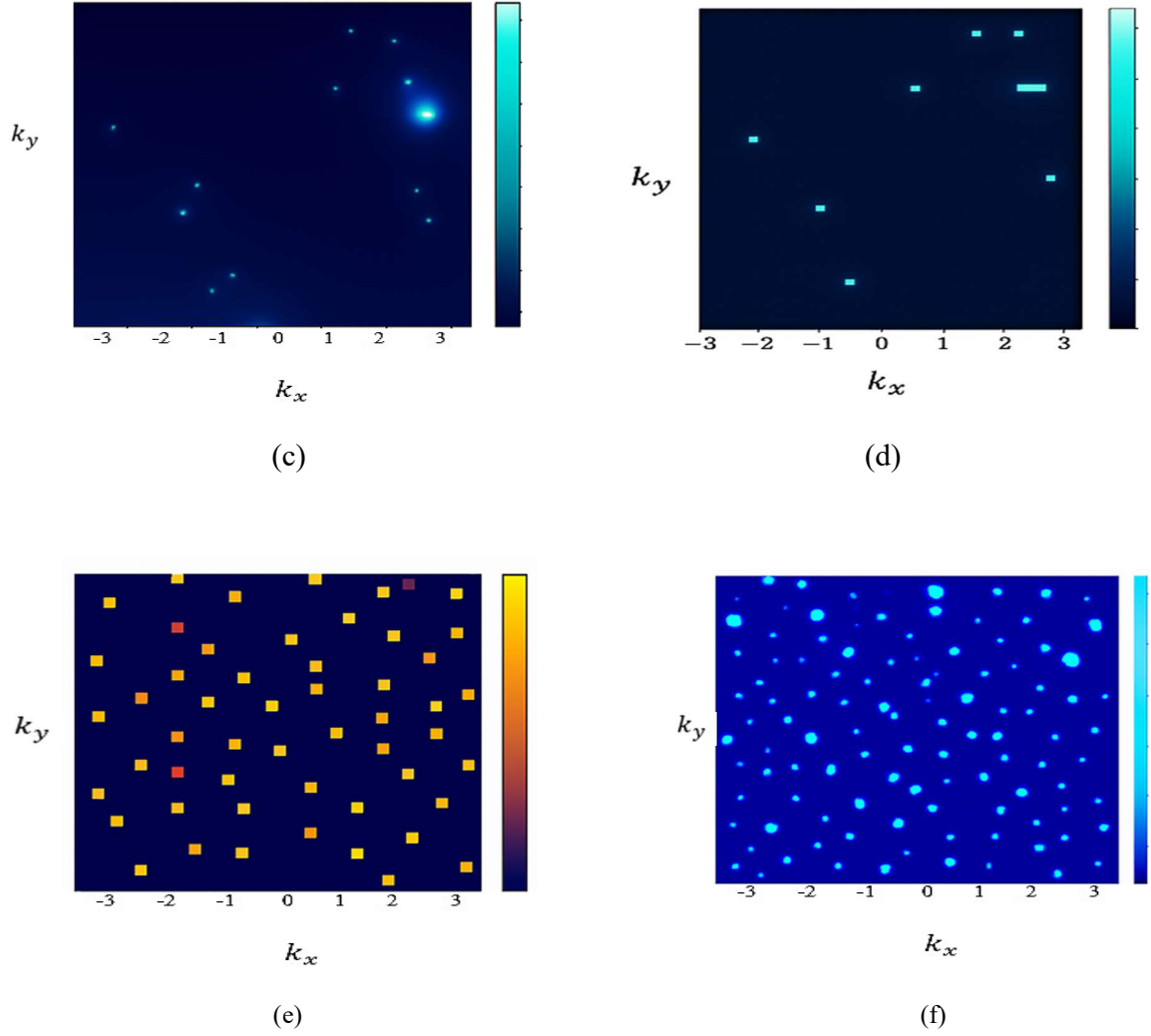

**Fig.5.** The 2D plots of the Berry curvature, in Figs. 5a, 5b, 5c, 5d, 5e, and 5f , as a function of $k_x$ and $k_y$ $\{k_x, k_y \in (-\pi, \pi)\}$ corresponding to the energy bands $E_1$, $E_2$, $E_3$, $E_4$, $E_5$, and $E_6$ in Fig.2b, respectively. The Chern number for the Berry curvature (Fig. 5a) of the band $E_3$ (Fig. 2b) is $C_1 \approx +1$, and that (Fig. 5b) for the band $E_4$ (Fig. 2b) is $C_2 \approx -1$. The rest of the bands (Berry curvature plots of which are shown in Fig. 5c, 5d, 5e, and 5f) have $C \approx 0$. The numerical value of the parameters used are $t = 1$, $t' = 0.86, \lambda_R = 0.8, J = 2.3$, and $\Delta_{CDW} = 0.14$. The phase ϕ of the complex NNN hopping ($t'$) is assumed to be momentum-dependent: ϕ(k) = (sin($k_x$) – sin($k_y$)). This introduces momentum-space winding, mimicking an orbital magnetic flux. In these figures, the color bar is comprehensively annotated, ranging from –5 at the bottom to +2 at the top, with intermediate tick marks clearly positioned at –4, –3, –2, –1, 0, and 1.

In Fig.5, we have the 2D plots of the Berry curvature, in Figs. a, b, c, d, e, and f, as a function of $k_x$ and $k_y \{k_x, k_y \in (-\pi, \pi)\}$ corresponding to the energy bands $E_1$, $E_2$, $E_3$, $E_4$, $E_5$, and $E_6$ in Fig.2b, respectively. The phase ϕ of the complex NNN hopping ($t'$) is assumed to be momentum-dependent: ϕ(k) = (sin($k_x$) – sin($k_y$)). This introduces momentum-space winding, mimicking an orbital magnetic flux. The Chern number for the Berry curvature (Fig. 5a) of the band $E_3$ (Fig. 2b) is $C_1 \approx +1$, and that (Fig. 5b) for the band $E_4$ (Fig. 2b) is $C_2 \approx -1$. The rest of the bands (BC plots of which are shown in Fig. c, d, e, and f) have $C \approx 0$. The numerical value of the parameters used are $t = 1$, $t' = 0.86, \lambda_R = 0.80, J = 2.3$, and $\Delta_{CDW} = 0.14$. With RSOC cranked up to 0.8 and exchange field strength boosted to 2.3, our KV$_3$Sb$_5$ system is now entering a regime of strong spin splitting and magnetic asymmetry. Combined with the momentum-dependent ϕ, this setup flattens bands and intensifies Berry curvature hotspots. It is important to note that the non-zero Berry curvature associated with bands $E_3$ and $E_4$ contributes to an intrinsic anomalous Hall effect, despite the total Chern number summing to zero. Given that the Fermi level currently intersects multiple bands, the system exhibits characteristics of a topological metal. However, if the Fermi level were tuned to lie precisely between $E_3$ and $E_4$, and a full band gap were opened, the system could transition into a Chern insulating phase. In such a scenario, a net chiral edge mode would emerge, contingent on which of the topological bands is occupied. As we notice here, the intensified RSOC and the exchange interaction J, in conjunction with the momentum-dependent phase (MDP) ϕ (ϕ(k) = (sin(k$_x$) – sin(k$_y$)), are essential to unravel that the system as a whole is no longer topologically inert under these conditions. The vital question here is whether MDPs in lattice models should ideally be rationalized beginning from the Peierls phase substitution (PPS); we note the following regarding this issue below.

In order to derive MDP contribution in a tight-binding model with NNN hopping, we consider the Hamiltonian term $[\sum_{r,\delta} t' e^{i\phi_\delta} c_r^\dagger c_{r+\delta} + \text{h. c.}]$, Here, $\delta$ symbol denotes the NNN displacement vectors, and $\phi_\delta$ are direction-dependent phases assigned to each hopping path. Upon Fourier transformation, this term becomes $\sum_\delta t' e^{i(k\cdot\delta + \phi_\delta)}$. To simulate a staggered flux pattern, the phases $\phi_\delta$ are chosen such that hopping along ± x direction acquires phases ± ϕ and similarly for y-direction. This antisymmetric assignment ensures that the net flux through each plaquette alternates in sign, mimicking a zero-net-flux configuration that breaks time-reversal symmetry locally. The imaginary part of the resulting sum captures the sine components of the dispersion, that is $Im \sum_\delta t' e^{i(k\cdot\delta + \phi_\delta)} \propto t'(\sin(k_x) - \sin(k_y))$. This is because the imaginary part of the exponential picks up the sine terms, and the alternating signs from $\phi_\delta$ encode the staggered flux. The question has been raised above should MDP in a lattice models be ideally rationalized starting from PPS, the standard method for incorporating orbital effects of magnetic fields. To address this, it's important to recall that in the presence of a real magnetic field, electrons acquire a Peierls phase—an additional complex factor in their hopping amplitude—proportional to the magnetic flux enclosed by their trajectory. This phase emerges naturally from a real-space gauge-theoretic treatment of electromagnetic fields. However, in systems engineered to have no net magnetic field, one can simulate similar effects by assigning complex phases to hopping terms such that the total phase accumulated around a closed loop mimics a magnetic flux. This construction, known as a staggered flux pattern, involves loops with alternating positive and negative fluxes. While the global flux cancels out, the local breaking of time-reversal symmetry gives rise to nontrivial Berry curvature. Thus, equating momentum-dependent hopping phases directly with the Peierls substitution introduces a conceptual mismatch. The Peierls substitution is grounded in physical gauge fields and real-space considerations, whereas MDPs often arise from synthetic or phenomenological models designed to emulate topological effects as in the Haldane model **[43,44]**. Although both approaches can yield similar topological signatures—such as nonzero Chern numbers—their physical origins and symmetry properties are fundamentally distinct.

**4. Future perspective and concluding remarks**

Given that our analysis reveals two occupied bands with opposite Chern numbers, a natural question arises: how can such a system support a quantized anomalous Hall effect (QAHE)? To address this, we propose a van der Waals heterostructure (vdWh) composed of a $KV_3Sb_5$ layer and a monolayer transition metal dichalcogenide (TMD), such as $MoS_2$ or $WSe_2$. The $KV_3Sb_5$ layer contributes six bands ( 6 by 6 Hamiltonian: $[H_{KV_3Sb_5}(\boldsymbol{k})]$ characterized by strong Berry curvature and potential topological features, while the TMD layer contributes six bands [6 by 6 Hamiltonian: $T(\boldsymbol{k})$] derived primarily from the $d$-orbitals of the transition metal, exhibiting strong spin-orbit coupling (SOC) and valley-dependent physics. In monolayer TMDs, the valence bands predominantly originate from the $d_{x^2-y^2}$ and $d_{xy}$ orbitals, while the conduction bands are mainly derived from the $d_{z^2}$ orbital. Strong SOC, particularly in tungsten-based compounds, leads to significant spin-splitting. At each valley (K and K′), this results in two spin-split valence bands and two spin-split conduction bands. Additional bands may arise from higher or lower energy d-orbitals or through substrate-induced hybridization. Thus, it is reasonable to identify six distinct bands in the TMD layer that exhibit strong SOC and valley-selective behavior. The feasibility of such a band structure within a vHs stems from favorable band alignment and interlayer coupling. The vdWh interface preserves the intrinsic electronic properties of each constituent layer while allowing for proximity-induced effects. Notably, the SOC from the TMD layer can influence the electronic states of $KV_3Sb_5$, and vice versa. Furthermore, the breaking of inversion and/or time-reversal symmetry at the interface can enhance Berry curvature, potentially enabling the emergence of QAHE when magnetic ordering is introduced. This heterostructure provides a 12-band basis, leading to a momentum-space Hamiltonian defined on a 12×12 matrix:

$$\begin{pmatrix} H_{KV_3Sb_5}(\boldsymbol{k}) & T(\boldsymbol{k}) \\ T^\dagger(\boldsymbol{k}) & H_{KV_3Sb_5}(\boldsymbol{k}) \end{pmatrix} \tag{10}$$

This framework sets the stage for exploring topological phases and spin-valley coupled transport phenomena in engineered quantum materials. Our upcoming work focuses on probing a van der Waals heterostructure comprising a TMD monolayer and $KV_3Sb_5$. Preliminary findings suggest that this hybrid system may support QAHE, offering exciting prospects for topological electronics. It is important to note in this context that although QAHE is intrinsically a two-dimensional phenomenon, the layered architecture and topological band characteristics of $KV_3Sb_5$ render it a compelling candidate—particularly within engineered environments as discussed above. Moreover, despite its bulk three-dimensional nature, $KV_3Sb_5$ can be exfoliated to few-layer or even monolayer forms, thereby enabling the potential realization of QAHE in quasi-2D regimes **[45]**.

We provided in this paper an explicit example of the Bloch Hamiltonian matrix in momentum space for a Kagome lattice of $KV_3Sb_5$ with one orbital, one layer, three sublattices (A, B, C), and two spin states (↑, ↓). It helps visualize the structure without being overly large. This setup gives the six-band(s) structure. We have calculated the Chern numbers for the bands. Our analysis reveals that the Nernst conductivity displays a striking non-monotonic temperature profile: it rises with increasing temperature, peaks prominently, and then diminishes at elevated temperatures due to thermal broadening, which suppresses Berry curvature contributions. Intriguingly, even modest shifts in chemical potential can induce pronounced changes in the Nernst response—amplifying its magnitude or inverting its sign—underscoring the system's acute sensitivity to carrier density. Within this landscape, we discern a tantalizing possibility for realizing the quantum anomalous Hall effect (QAHE), albeit through a configuration involving two bands with opposite Chern numbers. These findings form the conceptual core of the paper. It needs to reemphasize that despite its theoretical allure, QAHE remains experimentally elusive, hindered by both conceptual and material challenges. A key requirement is the opening of a magnetic-interaction–driven bulk gap at Dirac points—an effect not yet conclusively observed in $KV_3Sb_5$. Moreover, the realization of QAHE demands exceptionally clean systems, well-isolated bands, and precise tuning of the chemical potential. In practice, $KV_3Sb_5$ samples often contend with disorder and imperfect band alignment, which may obscure or suppress the desired topological signatures.